**NEMO: Neural Electro-Mechano-Optic Sensors for Multiplexed Neural Interfaces**

Andrew Cochran[1], Harshvardhan Gupta[1], Vishal Jain[1,2], Maysamreza Chamanzar[1,2], Gianluca Piazza[1].

1. Department Electrical and Computer Engineering, Carnegie Mellon University, Pittsburgh, USA 15213
2. Neuroscience Institute, Carnegie Mellon University, Pittsburgh, USA

**Abstract**

We introduce a novel electro-optomechanic neural sensor for realizing ultra-compact neural recording probes that can detect and relay electrophysiology signals from within neural tissue. This technology addresses outstanding challenges faced by existing neural recording technologies, including the resolution trade-off with signal-to-noise-ratio (SNR) due to the high impedances of small electrodes, and lingering stimulation artifacts. The sensor employs a highly miniaturized NEMS (nano-electromechanical systems) electrostatic transducer that modulates a silicon photonic microdisk resonator to convert electrical signals to an optical signal modulation. We have been able to achieve a limit of detection down to 110 microvolts, making the sensor sensitive enough to detect neural signals. This sensitive electro-optomechanic sensor directly detects electrophysiology signals and converts them to optomechanic modulation for effective transmission to outside the brain, which provides the unique potential for massive multiplexing of neural recordings. This design eliminates the need for bulky backend headstages that limit neural recording on awake free-roaming subjects. The ability of the device to record electrophysiological signals has been demonstrated using benchtop characterization and ex-vivo recordings from live neural tissue.

**Introduction**

Recording electrophysiological signals with high spatiotemporal resolution across multiple brain regions is essential for understanding how information is encoded and transformed within the central nervous system[1-4]. Needle-like implantable neural probes, equipped with an array of microscopic electrodes, offer the highest spatial and temporal resolution for recording signals from within the brain[5,6]. These electrodes, which are often much smaller than a single neuron, record the changes in electric potential due to transient variations in the electric field and ion concentrations within the neural tissue. Given the intricate connectivity in the brain, it is not only desirable to finely resolve neuronal activity in specific regions but also to do so simultaneously in different areas of the brain[4,7-9]. This necessitates multiple probes implants, which is very challenging with due to the size of state-of-the-art probes[10-15].

Despite significant advancements over the past decade in the scaling of neural probes, core design concepts have largely remained unchanged. Passive probes consist of micro electrode arrays connected to conductive wires along the probe's length, which end in a connector that attaches to a headstage. The headstage, a printed circuit board (PCB), has neural amplifiers, multiplexers, and digitizers that condition and transmit data over digital interfaces to a computer. This approach currently limits passive commercial probes to 128 channels per implant due to the large PCB fanout and the number of connections that need to be made to the headstage[16,17]. The headstage, weighing between 800 milligrams to 3 grams, is physically attached to the animal's skull using an enclosure and mount that can add several grams and bulk to the assembly[10-15]. This seriously hinders the ability to have multiple implanted probes, especially in experiments on smaller animals like rodents and small primates that are commonly used in neuroscience. Even a single implanted probe with its heavy backend connector, headstage, and tether can severely restrict the mobility of subjects, making multiple implants nearly impossible. There is a need to develop smaller and lighter backends that minimize the load on the subject.

In addition, conventional electrical recording systems face significant limitations in scaling down electrode size for smaller and denser arrays, primarily due to the relatively low input impedance (around 10 MΩ) of AC-coupled electronic neural amplifiers, which have a large input capacitance to block any DC offset. This necessitates larger recording electrodes, often coated with materials like PEDOT:PSS, to maintain low electrochemical impedance for signal capture[18-20]. However, the degradation of these coatings and glial scarring around electrode surfaces can substantially increase electrode impedance, negatively impacting neural recording quality[21-24]. Furthermore, a critical trade-off exists between electrode impedance and Signal-to-Noise Ratio (SNR). While smaller electrodes are desirable for enhanced spatial resolution and detecting specific cell types and signal propagation, their electrode-tissue interface impedance increases with decreasing size[20]. This increased impedance can become comparable to the recording amplifier's input impedance, limiting gain and deteriorating SNR. Similarly, neural amplifiers are constrained by a trade-off between gain/SNR and dynamic range, making them susceptible to stimulation artifacts[25, 26]; achieving higher gain (and SNR) restricts the amplifier to a small input dynamic range due to limitations in output swing range, device biasing, and operating voltages.

Another consequence of the large capacitance of neural amplifiers is their long discharge time. Experiments often require measuring the neuronal activity that is evoked in response to an applied electrical stimulation, and almost all closed-loop neurostimulation (CLS) paradigms involve stimulating and recording from the brain. Though relatively short pulses and small currents are used (100 μA – 1 mA, 100 μs – 5 ms), they can lead to potential swings large enough to saturate the neural amplifiers and charge up their input capacitors[25-27]. These are seen as stimulation artifacts in neural recordings. Though blanking circuits can be used to reduce these artifacts, the discharge is still a visible exponential decay, called the artifact tail, that can take several milliseconds to discharge and often occludes the desired electrophysiology signal. Techniques are used to fit exponentials to correct baselines and extract the "true" neural signal, but signals are often distorted and any activity during the blanking period is always lost[25, 26].

Recently, active neural probe designs have emerged that incorporate a portion of the headstage electronics like analog amplifiers, multiplexers, and digitizers on the probe itself. This approach aims to increase the channel count with greater integration embedded on the probe, amplifying the recorded neural signals using CMOS circuits and implementing active switching and low levels of multiplexing to reduce the number of interconnects[28]. This has allowed the number of recording electrodes to be greatly increased (up to about 1000 recording electrodes per shank), but only a subset are addressable over a few 100s of channels (i.e., Neuropixels[29]: 384, SiNAPS[30]: 512, NeuroSeeker[31]: 768). While this approach overcomes some of the interface related limitations to achieve higher channel counts, the size of the headstage and the overall implant remains bulky, due to the additional electronics that need to be added on the headstage PCB to control the on-probe circuitry (Figure 1). Having active circuits on the probe also leads to increased power consumption, and more importantly, increased heat generation in the brain, which is detrimental to neural tissue viability[32]. Moreover, challenges associated with low amplifier impedance and long discharge time constants remain[27].

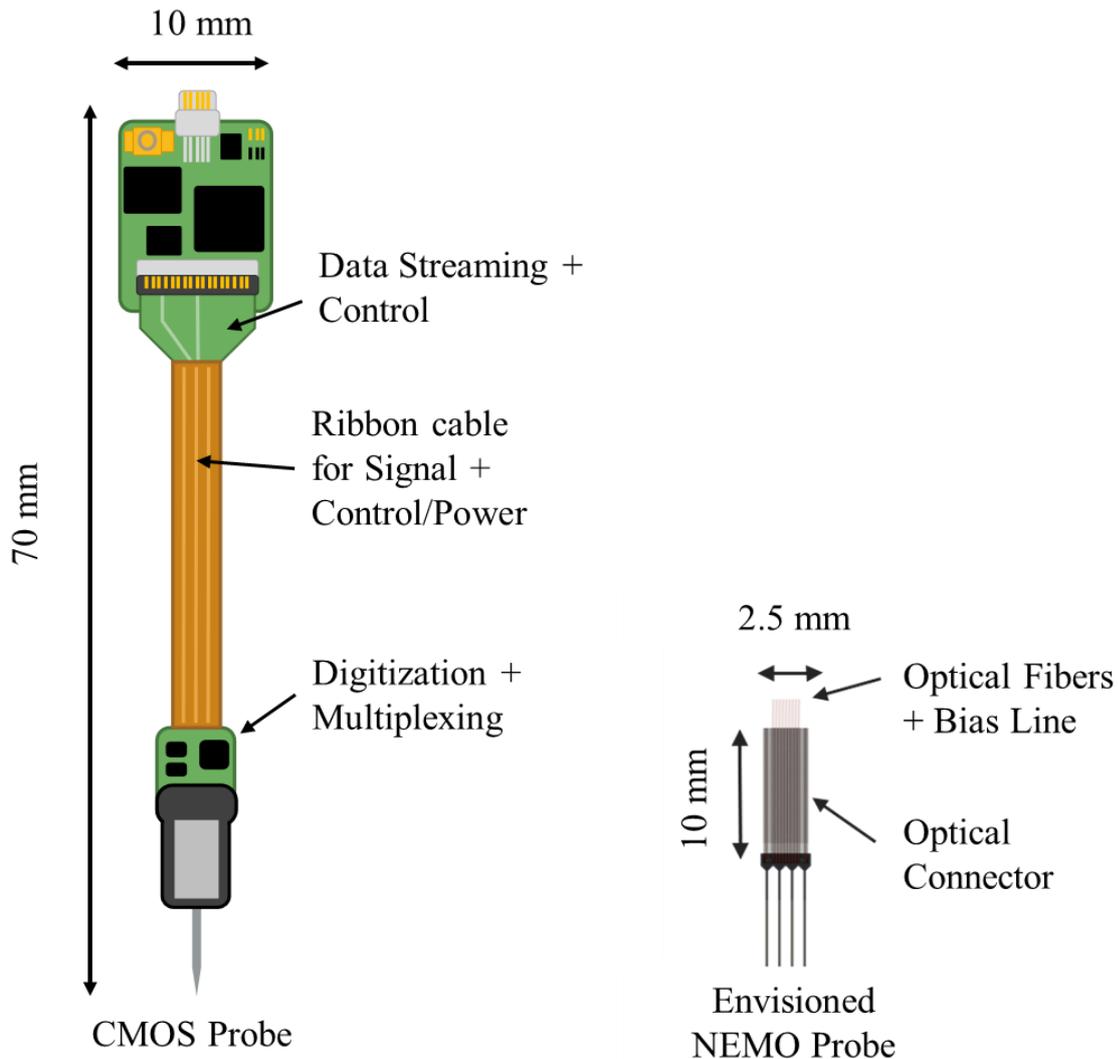

**Figure 1.** Dimensions of state-of-the-art neural probes. While the probe shank is very narrow (~ 100 μm) which minimizes tissue damage, the backend connectors and headstages (needed for fanout, digitization, and data transmission) are large. This makes use of multiple probes impractical, especially in chronic applications. The NEMO sensor is a multiplexor and encodes the neural signal optically allowing the signals from an array of sensor to be transmitted over an optical fiber without the need for electronics on the backend of the probe.

While CMOS-based neural probes have advanced the state-of-the-art with higher density electrodes, these techniques are reaching scalability limits. Even using smaller CMOS technology nodes will not fundamentally improve analog performance as gain and noise degrade[33]. To address these challenges, we look to a trend seen in telecommunications and semiconductor technologies where copper cables and interconnects are being replaced by optical fibers and co-packaged photonics[34]. Optical communications is the universal standard for ultra-high bandwidth communication where multiple electrical channels can be encoded onto THz frequency optical carriers and multiplexed over a single fiber. Furthermore, negligible optical signal attenuation and eliminating high-speed electrical I/O significantly reduces power consumption and is immune to electromagnetic interference. This enables data acquisition and digitization off-chip, without the need for bulky backends. Recent growth of photonic integrated circuits (PICs) has enabled miniaturization of these optical communication components in an integrated platform using the same fabrication processes as commercial CMOS foundries[35]. A neural probe leveraging optical communication through PICs, could provide a breakthrough in the scaling of high-density neural probes.

## Results

Here, we demonstrate our novel neural electro-mechano-optic (NEMO) sensors that are based on the modulation of a photonic microdisk resonator with a very sensitive NEMS (nano-electromechanical systems) actuator. The devices can achieve sensitivities that are several orders of magnitude higher than the state-of-the-art photonic modulators, allowing them to be employed for neural recording. The photonic microdisk resonator encodes the detected neural signal onto a distinct resonance wavelength, allowing potentially hundreds of devices to be coupled to the same bus waveguide.

In addition, the electro-mechanical sensor exhibits an extremely high input impedance (~48 GΩ), well beyond the limits of state-of-the-art neural amplifiers (i.e., Intan[40]: 13 MΩ, ADInstruments[41]: 100 MΩ). This decouples the effect of electrode-tissue impedance from the rest of the recording system, thus breaking the impedance-SNR trade-off, a longstanding limitation of existing neural interfaces. Therefore, this design facilitates scaling to small recording electrodes (with high impedances) to more effectively pick up single unit recordings, without affecting the SNR, and enable detection of different cell types from their corresponding spike waveforms[42,43]. Furthermore, the combination of a small electrode size, low input capacitance of the device (measured to be ~3.3 fF) and large dynamic range significantly reduces the stimulation artifact from several milliseconds to well below 1 ms ($\tau_{NEMO} = 0.073\ ms$). As a result, the stimulation artifact decays rapidly and will have a negligible overlap with the evoked response.

**Design**

To the best of our knowledge, there is no photonic sensor design reported in the literature that meets all the requirements for neural sensing and multiplexing. A sensor that can optically multiplex neural signals must have high sensitivity at low operating voltages with a sub-millivolt limit of detection, low power consumption for negligible heat generation, high input impedance, operating bandwidth > 30 kHz, and narrow spectral linewidth to multiplex multiple signals. The closest designs are photonic modulators, typically using thermo-optic or electro-optic transduction mechanisms, where the input voltage generates heat or an electric field that results in a change of refractive index. When built on a micro-disk resonator, this results in a change of resonance wavelengths which isolates the signal to a narrow wavelength linewidth. Using this technique, multiple modulators can be simultaneously coupled to a single waveguide[44,45].

These devices are typically used as high-speed electro-optic modulators or tunable filters rather than high sensitivity for low input voltages. However, if used as electrical neural sensors, each of these existing modulators have significant drawbacks. A thermo-optic modulator would not be able to detect the weak electrophysiology signals. Using a resistive heater to convert the electrophysiology voltage into thermal energy, a thermo-optic modulator, with the necessary input impedance > 1 MΩ, would result in negligible thermal power from the neuron ($P_{thermal} = V^2/R \approx 1 pW$) and would not cause any detectable modulation even using the most sensitive thermo-optic microring resonators (~ 1nm mW$^{-1}$)[36,37]. Existing electro-optic modulators transduce the electric field directly but typically have low sensitivities on the order of ~ 1 pm V$^{-1}$ or high optical losses which broadens the linewidth of the optical signal and limits multiplexing[38,46,47].

Recently, NEMS monolithically fabricated with photonics have been demonstrated with electro-optic sensitivities that are several orders of magnitude larger (up to ~ 1000 pm V$^{-1}$) . However, these devices require high operating voltages > 25 V that makes them vulnerable to noise and increased crosstalk, rendering them impractical for biomedical devices[39]. These devices also occupy large footprints or have broad linewidths that limit their usefulness for a multiplexed neural sensor array.

**Conceptual Design**

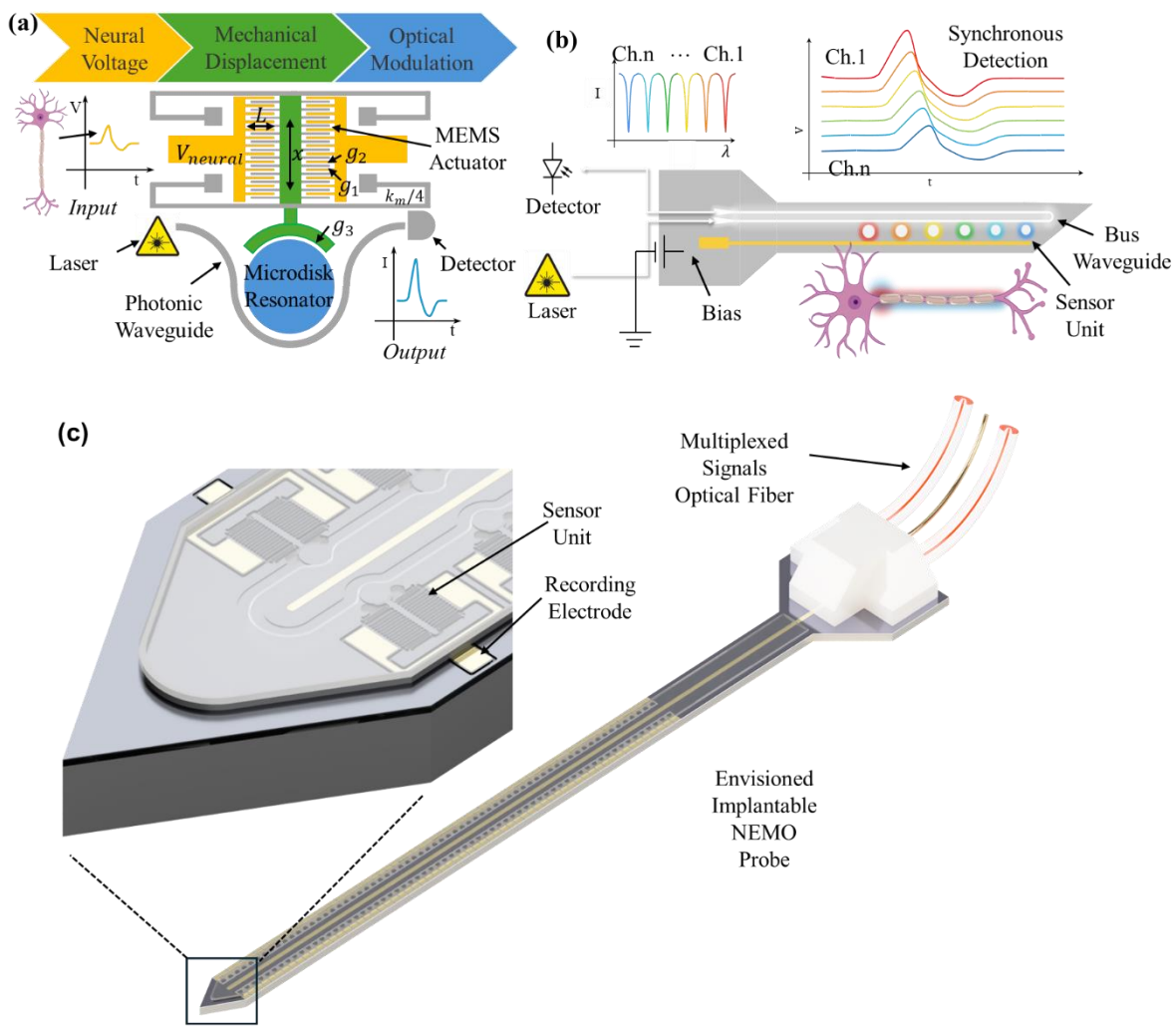

**Figure 2**. (a) Schematic of the NEMO unit cell transduction mechanism. Neuronal activity generates a voltage which is applied to the fixed fingers of the comb drive. This generates an attractive electrostatic force between the fixed and moving fingers that causes mechanical displacement of the comb drive. This mechanical displacement changes the resonance wavelength of the photonic disk. A laser is connected to the photonic waveguide which couples to the resonator and then to a photodetector. The input from the laser is modulated by the resonance wavelength change and neural signal is recovered at the output as the photodetector current. (b) Multiplexing of multiple sensors using a single waveguide by encoding each sensor at a different wavelength. (c) Schematic and enlarged view of NEMO sensors in an implantable neural probe where the backend connection is only a few optical fibers.

In this design, a NEMS actuator is first used to convert the electrophysiology voltage into a mechanical displacement, where the voltage generates an electrostatic force of a suspended comb drive. This motion is encoded optically as the actuator moves toward the photonic modulator causing a shift of its resonance wavelength (Figure 2a). A bus waveguide supplies light from a laser source and the photonic resonator filters out its respective resonance wavelength. A photodetector is then used to measure the transmitted light and reconstruct the electrophysiology signal (Figure 2b-c). By designing each photonic resonator to filter a different wavelength of light, multiple sensors can be connected using a signal bus waveguide in a wavelength division multiplexing (WDM) scheme. This technology would enable a lightweight backend, necessary for high density chronic recordings, where a single optical fiber could transmit the electrophysiology signals from multiple neural sensors, and just a few fibers could massively scale the number of recording channels to well beyond the state of the art (Figure 2c).

We have successfully designed novel electro-mechano-optical sensors that can detect electrical signals as small as 110 µVrms. By leveraging a gap closing actuator (GCA) design, we can scale the sensitivity of the device (> 150 pm V$^{-1}$) and the operating voltage (< 5 V), while maintaining low power consumption (< 1 µW), sufficient bandwidth (> 60 kHz), and high quality-factor (> 30 k, narrow spectral linewidth < 50 pm). This represents a 2 order-of-magnitude increase in sensitivity within a similar footprint for high-Q photonic NEMS resonators. The electro-optic transduction of the device is comprised of two components: the electro-mechanical sensor and the opto-mechanical transducer.

**Electro-mechanical design**

The electro-mechanical sensor has a sensitivity of $S_{mech} = dx/dV$ (where $x$ is displacement and $V$ is voltage), which depends on the electrostatic force, $F_e$, and the equivalent stiffness of the device, $k_m$. In contrast to conventional electrostatic comb drives, where the direction of displacement of the moving shuttle is along the length of the fingers, in the GCA, the direction of displacement is normal to the length of the fingers, where the gap between fingers changes. This results in an increased change in capacitance and electrostatic force compared to previous designs[38,40-43]. Since electrostatic forces are always attractive, we use an asymmetric gap arrangement to generate a larger force in the direction of actuation.

The electrostatic force, Fe, generated by the actuator is given by:

$$F_e = \frac{1}{2}\frac{\partial C}{\partial x}V^2 \approx \frac{NhL\epsilon_o}{2}\left(\frac{1}{(g_1 - x)^2} - \frac{1}{(g_2 + x)^2}\right)V^2, \quad (1)$$

where $N$ is the number of finger pairs, $h$ is the height of the fingers out of plane of the actuator, $L$ is the overlap length between fingers, $\epsilon_o$ is the dielectric constant of the medium, $g_1$ and $g_2$ are the finger gaps, $x$ is the deflection of the actuator, and $V$ is the potential difference between the two sets of fingers. By decreasing the gaps and width of the fingers, the density and electrostatic force increases.

The moving shuttle of the comb drive is suspended using 4 parallel silicon flexures that can be modeled as guided cantilever beams. The stiffness ($k_m$) of such a configuration can be calculated as:

$$k_m = \frac{12EI}{L_m^3} \text{ where, } I = \frac{hw^3}{12} \tag{2}$$

$L_m$ is the effective length of the beam, $E$ is the elastic modulus of the material (silicon), and $I$ is the second moment of inertia of the beam cross section. The moment of inertia can be calculated as a function of the width ($w$) and height ($h$) of the beam, as shown in Eq. (2). We used folded beams in our design to relieve intrinsic stress in the device layer and to reduce the footprint of the actuator (Figure S1).

These equations can be combined using force equilibrium to obtain the voltage-displacement relationship:

$$V \approx \sqrt{\frac{2k_m x}{N\epsilon_o hL}\left(\frac{1}{(g_1-x)^2} - \frac{1}{(g_2+x)^2}\right)^{-1}} \tag{3}$$

To further enhance the sensitivity of the device to detect a sub-millivolt signal, a DC bias voltage ($V = V_{bias} + V_{signal}$) can be used to increase the sensitivity as the force is non-linear with voltage (1). This bias enhancement is limited by the operating range of the device which approaches an instability point (pull-in) where the change in electrostatic force ($k_e$) exceeds the mechanical stiffness ($k_m$) of the device. In this design, a mechanical stopper (the photonic resonator) is used to prevent the comb drive fingers from colliding and device failure at pull-in:

$$k_e = \frac{dF_e}{dx} = -k_m \tag{4}$$

Which occurs at a fixed displacement $x_{pi}$ that is only a function of the gap ratios:

$$x_{pi} \approx \frac{\left((g_1-x_{pi})(g_2+x_{pi})^3 - (g_2+x_{pi})(g_1-x_{pi})^3\right)}{2(g_1-x_{pi})^3 + (g_2+x_{pi})^3} \tag{5).}$$

The proof-of-concept actuators were designed over a conservative area of 100 μm × 100 μm, to match the width of a neural probe, with $N$ = 22 finger pairs ($L$ = 10 μm × $w$ = 600 nm) to maximize the sensitivity within the area constraint, asymmetric gaps ($g_1$ = 500nm, $g_2$

= 1000 nm) sized so that the maximum travel distance ($x_{pi}$) would reach the radius of the photonic resonator where it is most sensitive, and a stiffness of ~0.5 N m$^{-1}$ ($L_m$ = 30 μm × $w_m$ = 500 nm) to increase sensitivity and reduce the operating voltage, using a $h$ = 220 nm standard photonics device layer thickness. The dimensions of the actuator were constrained to ensure the minimum features could be robustly fabricated and the actuator was rigid enough to avoid catastrophic failure during pull-in. The initial design was implemented in a finite element method (FEM) simulation to analyze the effect of fringe fields that increase the effective area and force generated by the actuator, and the bandwidth of the actuator which is limited by its mechanical resonance at ~ 60 kHz (Figure S1). The simulated electro-mechanical response was fit to Eq. 3 which shows an electro-mechanical sensitivity > 150 nm V$^{-1}$.

To demonstrate scalability of the device, the presented actuator was further miniaturized, while maintaining the same electromechanical sensitivity and operating voltage. By minimizing and removing over-dimensioned structures, folding and resizing beams to increase density, and decreasing the finger gap size the area is reduced by nearly a factor of 10 (Figure 3c). Similarly, adding an additional fold and reducing the pitch between folds of the flexures allows the stiffness ($k_m$) to remain constant while decreasing the lateral dimension by a factor 2. The overall dimensions of the device are reduced to 35 μm × 40 μm, approximately the size of a single neuron for high density integration.

**Opto-mechanical design**

To achieve high density multiplexing, the opto-mechanical transducer was designed to minimize losses in the photonic resonator. We use a fixed photonic disk resonator coupled to a common bus waveguide. A curved silicon waveguide segment (arc) interacts with the photonic disk resonator to achieve modulation. Since only a small portion of the light is coupled into the waveguide segment, the optical loss is minimal which allows the resonator to maintain a high quality factor and the modulation mechanism is dominated by a change in resonance wavelength ($\Delta\lambda_o$) due to local effective index change $\delta n_{eff}$:

$$\Delta\lambda_o(x) = \frac{\lambda_o}{2\pi n_{eff,0}} \int_{-\frac{\theta}{2}}^{\frac{\theta}{2}} \delta n_{eff}(\theta, \lambda, g)\, d\theta, \tag{6}$$

where $\lambda_o$ and $n_{eff,0}$ are the resonance wavelength and effective index at $x = 0$, $\theta$ is the angle over which the arc interacts with the resonator, $g = g_0 + x\cos(\theta)$ is the gap between the arc and the resonator that changes from the initial gap, $g_0$, due to the mechanical displacement $x$, and $\delta n_{eff}$ is the effective index change due to resonance mode interacting with the arc as function of its position. This allows the mechanical sensitivity to be transduced into the optical domain as a change in resonance wavelength of the opto-mechanical structure.

The refractive index change ($\delta n_{eff}$) can be improved through the actuator design by increasing the arc interaction angle ($\sim\theta$), reducing the initial gap ($\sim 1/g_0$) where the electric field intensity and sensitivity of the optical mode is highest, or increasing the refractive index of the arc by introducing additional material to increase the contrast compared to the air cladding ($n_{arc} \neq n_{air}$). There are practical considerations in the design, including: the arc angle is limited by the need to connect to the bus waveguide, reducing the initial gap or introducing additional materials could decrease the quality factor of the device, and fabrication challenges associated with small features, long suspended structures, and additional materials. For the initial design, we use a simple silicon arc with an angle of 90-degrees and 150 nm gap to maximize the quality factor and manufacturability.

The refractive index change can also be enhanced by the photonic resonator Purcell factor. For a cavity with quality factor ($Q$), mode volume ($v_m$), and resonance wavelength ($\lambda_o$), the sensitivity is increased by the Purcell factor ($f$):

$$f = \frac{3Q\lambda_o^3}{4\pi^2 v_m}, \tag{7}$$

where the mode volume is proportional to the radius, ($v_m \approx 2\pi R t b$, here $t$ and $b$ are the vertical and lateral extent of the electric field) such that the sensitivity of the resonator scales with ($f \propto Q/R$) [52,53]. Thus, by reducing the size of the resonator, the sensitivity can be enhanced. When the radius of the resonator is significantly reduced ($R$ < 5 μm), bending losses within the resonator will increase which can degrade the $Q$ and limit further enhancement[52].

To detect this change in resonance wavelength, the resonator is coupled to a bus waveguide. Using coupled mode theory, the transmission function ($T$) of the photonic resonator is given by the equation:

$$T(\lambda) = \left| \frac{2j(\lambda_o/\lambda - 1) + 1/Q_o - 1/Q_c}{2j(\lambda_o/\lambda - 1) + 1/Q_o + 1/Q_c} \right|^2, \qquad (8)$$

where $\lambda, Q_o, Q_c$ are the wavelength of operation, intrinsic quality factor of the resonator, and coupling quality factor between the waveguide and resonator, respectively. This results in a device that is sensitive to changes in resonance wavelength ($\lambda_o$) and whose sensitivity is enhanced by the loaded quality factor $1/Q = 1/Q_o + 1/Q_c$. When the device is critically coupled ($Q_o = Q_c$), the transmission drops to zero at resonance, creating a high-extinction, narrow-linewidth response that can be leveraged to isolate the modulated signal. By probing with a laser near the resonance wavelength of the device, the change in transmission can be detected using a photodetector and an amplifier. A demultiplexing scheme is used to record the sensed signal.

To multiplex multiple sensors on a single bus waveguide, each sensor can be tuned to a different resonance wavelength by varying the radius, resulting in a unique resonance wavelength. However, these resonances have a periodic resonance condition ($m$) which limits the available linewidth to the free spectral range ($FSR$),

$$FSR \approx \frac{\lambda_o^2}{2\pi n_{eff} R}, \qquad (9)$$

where reducing the radius of the resonator increases the available linewidth. The theoretical bandwidth-limited multiplexing ratio ($M$) is determined by the number of sensors that can be multiplexed within a single FSR using sensor linewidth ($\delta\lambda = \lambda_o/Q$):

$$M \approx \frac{FSR}{\delta\lambda} \approx \frac{\lambda_o Q}{2\pi n_{eff} R} . \quad (10)$$

In practice, the multiplexing of photonic resonators is often limited by the ability to precisely control the resonance wavelength of the resonator based on the change in radius ($\Delta R$). Since the FSR and wavelength change ($\Delta\lambda$) are both inversely proportional to the radius ($1/R$), this multiplexing ratio is independent of radius given sufficiently high $Q$:

$$M \approx \frac{FSR}{\Delta\lambda} \approx \frac{\lambda}{2\pi n_{eff} \Delta R}, \quad \text{if } Q > R/\Delta R. \quad (11)$$

The proof-of-concept photonic structures were designed with a two different radii of 10 μm and 25 μm with 550nm wide bus waveguide supporting a single TE-like mode using a 90-degree pulley coupling configuration with a 100 to 300 nm gap to suppress spurious modes[54]. The perturbing arc has a width of 500 nm with an initial gap of 150 nm with a 90-degree angle. The design was implemented in a finite element method (FEM) simulation with a reduced radius to show the feasibility of the design, while achieving a reasonable computational speed and convergence (Figure S2). Simulations show an increase in resonance wavelength as the gap decreases, which is expected as the silicon arc ($n \approx 3.48$) increases the refractive index near the resonator compared to the air cladding ($n = 1$).

The theoretical multiplexing limit is analyzed using Equation 10 and 11. This shows that for a relative fabrication tolerance of 1 nm, ~ 100 channels can be transmitted over a single waveguide and would only require a $Q$ of ~ 5 k. If the tolerance can be improved to 0.1 nm, > 1000 can be multiplexed with a minimum $Q$ of ~ 50 k, which is reasonable given previous demonstrations of $Q > 200$k for 5 μm resonators[55]. 3D finite element method eigenmode simulations of a 5 μm radius resonator with the arc modulator showed a $Q > 300$k indicating that the modulator design does not introduce significant optical losses that would limit the multiplexing capability.

**Considerations for Neural Recording**

Beyond achieving high sensitivity to detect small electrophysiology signals, the sensor must also address system-level considerations including input impedance, power consumption, and suppression of artifacts from electrical stimulation commonly used in neuroscience experiments. Achieving a high input impedance is critical to neural amplifier performance to ensure that the gain of the system does not degrade due to the recording electrode impedance. For the NEMO sensor, the input port is the voltage applied across the electrostatic actuator. The device forms an air gap capacitor, defined by the overlap of the interdigitated fingers. From finite element simulations, the capacitance (C) is < 3.3 fF, which results in a device limited input impedance of ~ 48 GΩ at 1 kHz, several orders of magnitude larger than state-of-the-art neural amplifiers[40,41]. For the benchtop characterization in this work, the parasitic capacitance from the probe pads limits the impedance to ~0.3 GΩ, but in the packaged device would enable a impedance > 28 GΩ only limited by the parastic capacitance of the recording electrode **(**Table S1**)**.

      This capacitance also eliminates static power consumption of the actuator which consumes negligible dynamic power due to sub-nanometer actuation of device by the neural signal. The device power consumption is thus only limited to the optical power needed to detect the changes of the photonic resonance. In this work, an input power at the resonator of < 10 μW was needed, which was predominantly limited by losses of the optical grating couplers. In the simplest detection paradigm, where the laser is continuously set near the resonance wavelength to monitor the sensor, most of the power is dissipated by the photonic resonator (< 10 μW) which is still comparable or better than state of the art active neural probes[29,30]. In practice, each sensor only needs to be probed with the laser at the sampling rate as in a swept source configuration where the laser wavelength is continuously swept over the

entire wavelength range to probe multiple sensors. This would dramatically decrease the power consumption of each channel to < 25 nW as the resonance wavelength is scanned periodically ($P_{cons} \approx (P_{in} = 10\ \mu W) \times (\%\ of\ time\ at\ resonance \approx \delta\lambda\ /\ scan\ range = 50\ pm/80\ nm) \approx 25\ nW$).

The small capacitance of the device also allows charges to be dissipated quickly which can eliminate interference of stimulation artifacts with the electrophysiological response. In conventional amplifiers with much larger input capacitances (~ 10 pF) and saturation of the small dynamic range, the stimulation artifact persists for several milliseconds which overlaps and distorts the evoked response. Here, the capacitance of the device is nearly 4 orders of magnitude smaller (even with the large probe pads for benchtop testing it is nearly 2 orders of magnitude smaller) such that the device discharges charges rapidly with a time constant ($\tau \propto C$) that is negligible compared to state-of-the-art neural probes.

**Device Demonstration**

The devices were fabricated on standard silicon-on-insulator (SOI) photonic wafers (SOITEC) with a 220 nm thick device layer (Figure 3).

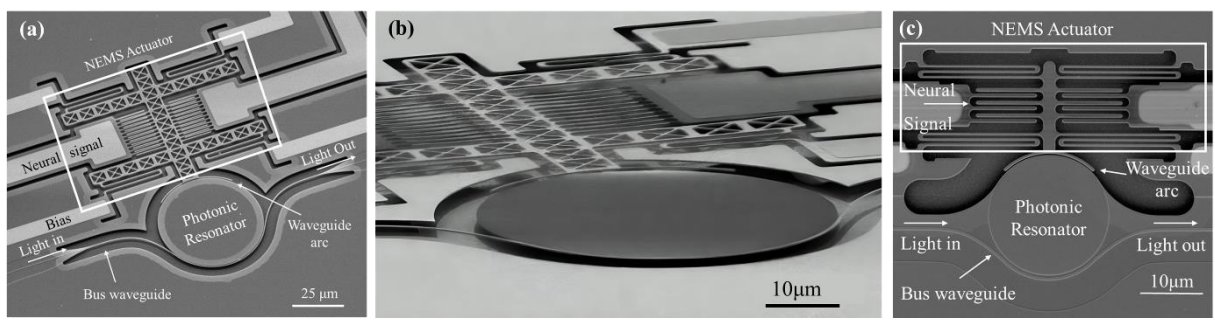

**Figure 3.** SEM images of the fabricated device: (a) The electrical signal is applied across the NEMS actuator terminals which causes a change in the resonance wavelength of the microdisk. Light from the laser passes through the waveguide which couples to the microdisk resonator and then connects to a photodetector to decode the input signal. (b) Tilted view of the sensor showing the waveguide and NEMS actuator suspended around the microdisk. (c)

Miniaturized sensor with a footprint of 35 μm × 40 μm which is ideal for scaling into a dense sensor array.

The photonic devices were characterized using a custom-built probe station and a tunable laser source (Santec TSL-510A) in the range of 1500 – 1600 nm. Light from the laser was carried over polarization maintaining optic fibers and coupled into the chip using angled V-groove arrays (VGAs) with an 8° facet angle. Electrical probing of the device was performed using GSG probes connected to a function generator.

**Sensor Characterization**

The optical transmission spectra of the photonic devices were obtained by sweeping the wavelength of the tunable laser while monitoring the output of the device using an amplified InGaAs photodetector (Thorlabs PDA10CS2). The output voltage of the photodetector was recorded using a data acquisition system (National Instruments NI PCIe-6374). Figure 4a shows the typical transmission spectrum of a device. The overall transmission spectral shape is dominated by the spectral response of the grating couplers. Sharp resonance dips, separated by the free spectral range of 4.3 nm, correspond to the resonance modes of the microresonator. The average quality factor of the fabricated photonic resonators was ~ 30,000 for the 25 μm radius resonators and ~ 10,000 for the 10 μm resonators, limited by the roughness of the etched resonator sidewalls.

    Figure 4b-c shows one of these resonance peaks with a DC voltage applied to the electrostatic actuator causing a shift in the resonance wavelength. Due to the non-linearity of electromechanical and optomechanical transduction, the shift in resonance wavelength increases dramatically at higher voltages. This shift continues until the electrostatic actuator pushes the arc into contact with the resonator which acts as a mechanical stopper. The gaps between the fingers are intentionally designed to be wider than the gap between the

waveguide arc and the photonic disk resonator to ensure that the fingers of the actuator do not collapse together during the pull-in process and prevent irreversible stiction. When the applied voltage is reduced below the pull-in voltage, the arc of the actuator is released, allowing the modulator to resume its functionality.

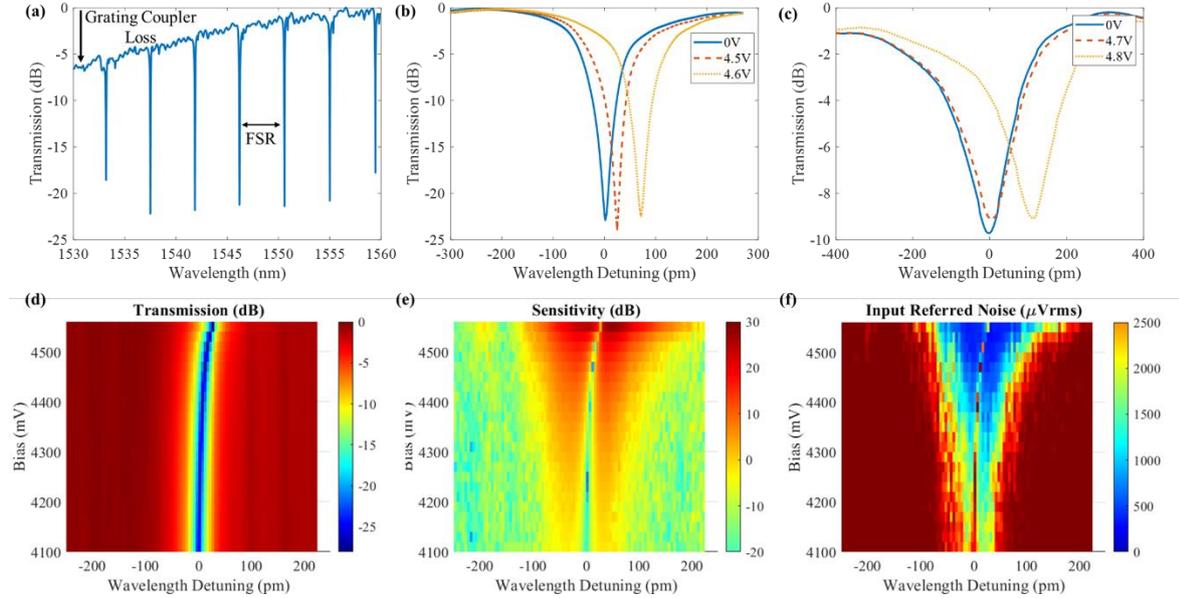

**Figure 4.** Optical transmission spectra of a device - (a) showing the expected FSR of ~ 4.3 nm and suppression of spurious modes using pulley coupling (b) resonance wavelength shift with applied DC bias for the device used in the remaining experiments, (c) miniaturized device showing similar sensitivity, but with a reduced Q ~ 10k due to the smaller resonator radius. (d) Transmission around a resonance mode, (e) the measured overall system sensitivity around the mode, and (f) input referred noise of sensor, all as a function of the bias voltage and wavelength detuning.

Sensing of smaller signals is achieved by adding a DC voltage (~ 4.5 V) to bias the device (like biasing a transistor) near pull-in where its electro-mechanical sensitivity (~ 550 nm V$^{-1}$) is maximized and the waveguide arc is moved close to the photonic disk. By sweeping the bias voltage and laser wavelength near the resonant peak, the maps in Figure 4 were obtained. The encoded optical signal is confined to the region near the resonance peaks leaving the rest of the optical spectrum unchanged.

Figure 4e shows the sensitivity of the sensor system: the ratio of the photodetector voltage to the small signal input voltage. As expected, the gain increases with increasing DC bias, reaching its maximum just before the actuator is pulled in. The gain as a function of wavelength follows the derivative of the Lorentzian resonance shape which is 0 at the resonance wavelength and maximized at the -1.25 dB bandwidth on either side of the resonance mode. Currently, the noise floor of the system is limited by the external I/O connections, predominately the laser, which accounts for nearly all of the 110 µVrms noise. To reduce the input-referred noise, the sensitivity of the device can be improved. The sensitivity is primarily limited by the quality factor of the photonic resonator. By increasing the quality factor of the devices to a very feasible value of 150 k (previous work has shown even a *Q* of 420 k is possible in a similar 10 µm radius resonator[55]),  the input-referred noise is scaled proportionally to be only 7.5 µVrms, effectively matching the capabilities of CMOS probes.

These plots also show how the gain is confined around the photonic resonance within ±100 pm (12.5 GHz) of the optical mode. With an FSR of 4.3 nm, this translates to 86 channels on a single waveguide. This number can be significantly increased by reducing the size of the resonator, thereby increasing the FSR, and by using other spectrum division techniques that allow the entire available range of the laser source to be utilized.

To verify the performance of the sensor, a 1 mV sinusoid test signal was applied to the device and the modulated optical signal at the photodetector was recorded to calculate the SNR (16dB) and input referred noise (110 µVrms) (Figure S3a). The bandwidth of the sensor is limited by the mechanical resonance of the actuator that can be seen as a small peak in the frequency response (Figure S3b) at ~ 60 kHz. The quality factor of the mechanical resonance is low (< 5) owing to air damping around the small and light structure. This reduces ringing and makes it robust to high frequency content from stimulation pulse artifacts (gain of the system is flat to less than ±1 dB up to 80 kHz).

Based on these benchtop characterizations, the fabricated device can be used to detect sub-millivolt signals with a noise floor of 110 µVrms, which is dominated by laser noise. This is achieved using a low bias voltage of 4.5 V, a sensitivity of > 150 pm V$^{-1}$, a bandwidth of 60 kHz, and spectral linewidth of 50 pm using optical input power of 10 µW. Therefore, our fabricated NEMO sensors provide a suitable solution to record electrophysiological neural signals and encode them into the optical domain to address the communication bottleneck of existing neural probes.

**Ex-vivo electrophysiology recording with NEMO unit cell:**

To further demonstrate the ability of our sensor unit cells to record real neural signals, we deployed our NEMO sensors to detect electrophysiology signals in an ex-vivo model with mouse brain slices (Figure 5). We used electrical stimulation to evoke electrophysiology signals that were recorded both by the NEMO sensor and also a conventional neural amplifier (as a reference). This stimulation-recording experimental paradigm clearly highlights the device dynamic range and small device capacitance that helps with suppression of the stimulation artifacts (Figure 7d).

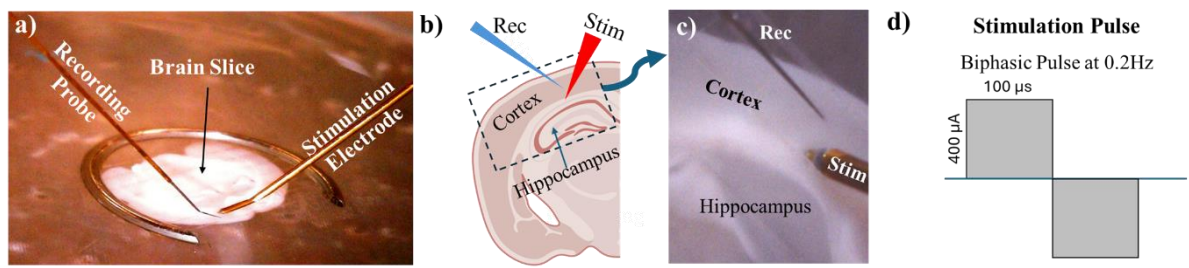

**Figure 5.** Experimental setup of the ex-vivo electrophysiology experiment. (a) Mouse brain slice with stimulation and recording electrodes placed on the cortex. (b) Schematic showing

electrode placement locations. (c) Microscope image of the stimulation and recording electrode placement in cortical layer 5 and 2/3. (d) Stimulation pulse.

First, we demonstrated the ability of NEMO sensor in the somatosensory cortex for recording neural response because it provides a well-characterized laminar structure with predictable anterograde connections (e.g., Layer 5 to Layer 2/3). Cortical circuits are widely studied in electrophysiology, making it easier to compare results with other similar approaches and validate the NEMO sensor's performance. In our experiments, stimulation was performed in layer 5 of the somatosensory cortex using an anodic-leading biphasic pulse with 100 μs pulse width at 0.2 Hz and the evoked response was recorded in Layer 2/3 using a tungsten probe connected to the NEMO sensor and neural amplifier.

With a stimulation current of 400 μA, the NEMO sensor successfully recorded the evoked neural response, which closely matched the ground truth measurement obtained from the conventional neural amplifier. The recorded signals exhibited strong agreement in both response magnitude and temporal onset, confirming the sensor's accuracy. To further refine the data, the trial-averaged neural response (n = 100) was processed using a 60 Hz notch filter, effectively removing the power line interference. This filtering was applied to both the electrical recordings from the neural amplifier and the optical recordings from the NEMO sensor, as illustrated in Figure 6c. Due to the improved dynamic range of our NEMO sensor (> 50 mVpp), the stimulation pulse is recorded without saturating the device, while the conventional neural amplifier recording shows significant distortion as the artifact exceeds the dynamic range (~10 mVpp). To verify that the recorded signals were of neural origin, we performed a control experiment in which the polarity of the stimulation pulse was inverted (cathodic-leading). As expected, this resulted in an inverted stimulation artifact, while the extracellular neural response retained its original polarity, further confirming that the detected signal was not an artifact (Figure 6d-e). Additionally, both the recording and stimulation electrodes were retracted from the brain slice while still submerged in the aCSF solution. As anticipated, only the stimulation artifact remained, with no detectable neural response.

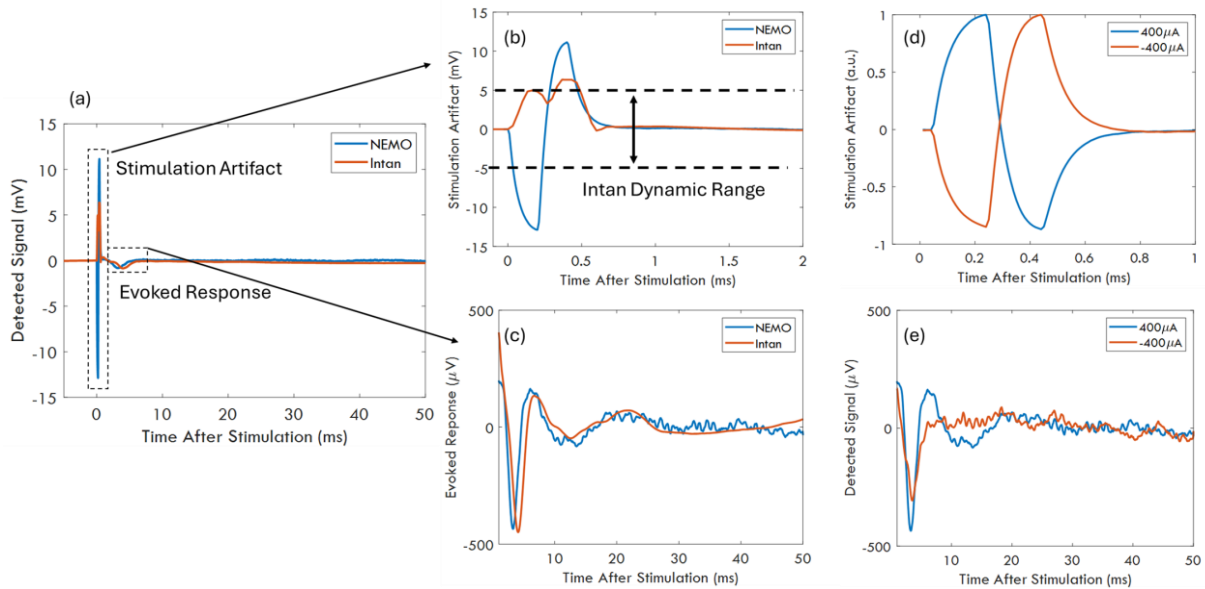

**Figure 6.** Cortex recordings: (a) Full recording of stimulation and evoked response from both the NEMO sensor and conventional neural amplifier. (b) Stimulation artifact recorded by the NEMO sensor shows a smooth RC response that follows the stimulation artifact. The neural amplifier recording (Intan) has significant distortion and non-linearities due to the small dynamic range of the amplifier needed to achieve large gain. (c) Recording of the neural signal from amplifier and NEMO shows good temporal and waveform agreement (trial averaged, n = 100). (d) Recordings of the stimulation artifact and (e) neural response with opposite stimulation polarity. The stimulation artifact becomes inverted, while the neural response polarity remains the same. This indicates that the recording signal is a neural response and not an additional artifact of the stimulation.

After establishing the sensor's performance in the cortex, we extended our experiments to the hippocampus, which presents substantially greater experimental and physiological complexity. Hippocampus is characterized by dense recurrent connectivity, prominent population synchrony, and strong synaptic plasticity mechanisms even in slice preparations[58,59]. Hippocampal recordings impose additional challenges, including increased tissue heterogeneity, signal attenuation, complex population dynamics driven by recurrent excitation and inhibition pathways[60,61]. The stimulation was applied to the Schaffer collateral layer, and the evoked response was recorded in the CA1 region, following the well-established synaptic pathway of this circuit. Here, we chose an electrode placement where significant lingering stimulation artifacts were observed with the conventional neural amplifier recording to showcase the NEMO sensor's ability to suppress these artifacts under the same conditions. The NEMO sensor's recordings showed strong agreement with the electrical recordings from the neural amplifier, confirming the accuracy of both modalities (Figure 7). These results show the reliability of the NEMO sensor in accurately capturing hippocampal neural activity, while effectively minimizing stimulation artifacts.

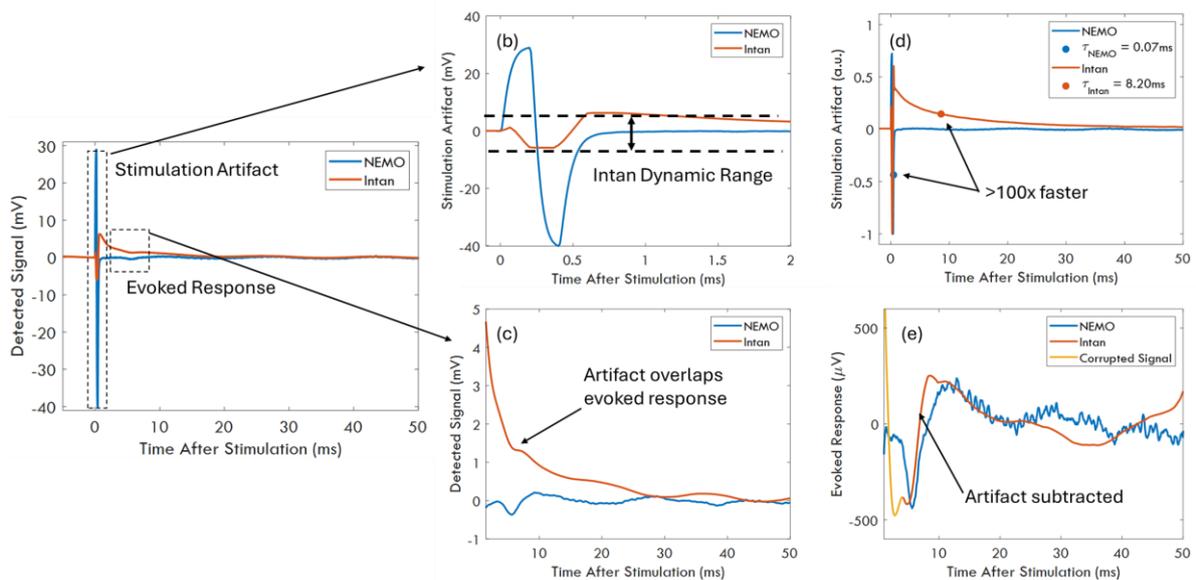

**Figure 7.** Hippocampus recordings: (a) Full recording of stimulation and evoked response from both the NEMO sensor and conventional neural amplifier. (b) Enlarged view of the stimulation artifact showing that the limited dynamic range of conventional neural amplifiers leads to saturation and distortion. (c) Enlarged view of the recorded signal where the artifact from the conventional amplifier overlaps with the evoked response due to the slow discharge of the amplifier capacitance. The NEMO sensor rapidly discharges so that the neural response is clearly visible. (d) Recording of the stimulation artifact without evoked response. Stimulation artifact takes much longer to decay ($\tau\_intan > 100\tau\_NEMO$) with neural amplifier compared to NEMO (recording without neural activity). (e) Recording of the neural signal from amplifier and NEMO shows good temporal and magnitude agreement, but the neural amplifier has additional features due to the stimulation artifact and noise that are too slow and large to be neural signals (trial averaged, n = 100). Fitting and subtraction of the decay of the stimulation artifact creates additional features not generated by the neural signal. This makes deciphering between neural signals and artifacts especially challenging.

This experiment highlights a major challenge of conventional neural amplifiers in recording evoked neural responses, the saturation of the amplifier, caused by strong stimulation artifacts. Due to the relatively large input capacitance of the conventional neural amplifier (12 pF), electrical stimulation results in a substantial artifact as the injected current charges and discharges the amplifier's input capacitance. As shown in Figure 7a-c, this artifact significantly exceeds the dynamic range of the conventional neural amplifier causing saturation and distortion that decays slowly and overlaps with the evoked response. A common approach to address this challenge is template subtraction, in which an exponentially decaying artifact signal is fit to the recorded signal. The fitted signal is then subtracted from the recorded signal to remove the artifact. However, the response of the amplifier after saturation is non-linear and may not be accurately modeled by an exponential fit. This can result in inaccuracies in the extracted evoked response that may be misinterpreted as electrophysiology signals (Fig. 7e).

To characterize the performance of the NEMO sensor in rapidly discharging the stimulation artifact, a stimulation current pulse was applied near the recording electrode and the time constant of the stimulation artifact decay ($\tau \propto C$, where $C$ is capacitance of the electrode-amplifier system) was measured using the conventional neural amplifier and the NEMO sensor. The NEMO sensor effectively mitigates this stimulation artifact issue (Figure 7d) by rapidly discharging the current ($\tau_{NEMO}$ = 0.073 ms) which is 2 orders of magnitude faster than the conventional neural amplifier ($\tau_{Intan}$ = 8.20 ms). This prevents amplifier saturation and enables accurate recording of the evoked neural response without the need for blanking, template subtraction, or reducing stimulation intensity—all of which can distort or obscure fast-occurring neural signals. Since the same electrode interface was used in both recordings representing the main resistive component, the capacitance of the NEMO system can be estimated, using a first-order RC circuit model, as $C_{NEMO} \approx C_{Intan} (\tau_{NEMO}/\tau_{Intan})$ ~ 100 fF (which shows reasonable agreement with the calculated parasitic capacitance) corresponding to an ultra-high input impedance $Z_{in} \approx 1/(2\pi f C_{NEMO})$ ~ 1 GΩ at 1 kHz. It is important to note that this impedance is currently dominated by the parasitic capacitance in the benchtop testing setup and can be drastically improved to be close to the intrinsic device capacitance (~3.3 fF from comb drive) in a packaged device with an integrated recording electrode (~2 fF of parasitic capacitance).

These results demonstrate the ability of the NEMO unit cell to record electrophysiology signals in the presence of stimulation artifacts. The NEMO sensor's recordings exhibit strong agreement with the ground truth electrical measurements of local field potentials (LFPs), confirming its accuracy in capturing neural activity. With a sufficient bandwidth of 60 kHz and a signal-to-noise ratio (SNR) of 16 dB at 1 mV, the sensor is highly sensitive and capable of detecting weaker, localized neural spikes. The sensor also significantly reduces stimulation artifact decay time by 2 orders of magnitude (~ 10 ms to ~ 0.1 ms). The rapid artifact suppression ensures that the evoked neural response remains clearly distinguishable, even in the presence of parasitic capacitance (Figure 7d). This unique capability is particularly advantageous when paired with smaller recording electrodes, further enhancing its utility for high-resolution electrophysiological recording.

**Discussion**

The demonstrated high sensitivity optomechanical neural sensor can detect and optically encode electrophysiology signals thanks to the several orders of magnitude improvement of the sensitivity compared to state-of-the-art photonic NEMS devices. One key advantage of this device is that it is an intrinsic multiplexor, which can greatly reduce the number of interconnects and backend size without the need for additional circuitry. Tuning the resonance wavelength of the sensors with different radii, enables an array of NEMO sensors to be transmitted on a single waveguide and optical fiber. The main challenges of scaling the device into a larger array are the sensitivity (to reduce input referred noise) and spectral density. Both metrics are currently limited by the quality factor of the device which is well below theoretical and previously demonstrated resonators with $Q > 420,000$ even for 10 µm radius devices[55]. Methods for improving the quality factor of photonic resonators are well established with standard CMOS fabrication processes to optimize line edge roughness of the lithography, reactive ion etching, and annealing techniques[55-57].

The ultra-low capacitance (~ 3 fF) interface of the sensor enables a new paradigm in electrophysiology recording where the long-standing limitations of electrode impedance and stimulation artifacts seen with conventional neural amplifiers are effectively eliminated. Even in the unpackaged device where parasitic capacitances dominate, the sensor drastically improves charge dissipation from current stimulation by two orders of magnitude, compared to a conventional neural amplifier[40,41], and distortion that compromises accurate recording of the evoked response[25,26]. The resulting input impedance can be two orders of magnitude higher than state of the art neural amplifiers which circumvents the requirement for stringent electrode impedance design and facilitates exploration of ultra-compact electrodes capable of cell identification[42,43]. The use of passive photonics and air-gap electrostatic actuators with no static power consumption provides a unique advantage for power and temperature sensitive applications, like electrophysiology recordings, compared to CMOS approaches that rely on active switching and amplifiers[29-31].

The miniaturization of the sensor into a compact footprint of ~ 35 μm × 40 μm enables implementation into a high-density implantable probe array. The narrow linewidth of the optically encoded photonic resonator is intrinsically multiplexed allowing an array of sensors to be transmitted over a single interconnect. This represents a paradigm shift from current active CMOS approaches which rely on bulky backend headstages and many interconnects for fanout, amplification, and multiplexing – impediments that severely constrain the practical use of multiple implantable probes.

**Materials and methods**

*Device fabrication:* The NEMO sensors were fabricated on SOI wafers with a 220 nm device silicon layer and a 2 μm buried oxide (BOX) layer. The NEMS and photonics structures were patterned using a 100 kV E-beam lithography system (Elionix) in CSAR AR-P6200.18 resist. The silicon device layer was micromachined using inductively coupled plasma reactive ion etching (ICP RIE, PlasmaTherm Versaline) using chlorine gas and released using a low-pressure vapor hydrofluoric acid etch. Two silicon etch steps were used - one 100 nm partial etch to define the grating couplers and the ridge waveguides, and a full 220 nm etch step to define the GCA, photonic resonator, arc modulator, and suspended waveguides. Details of the fabrication process flow are discussed in the Supplementary Information.

*Electrophysiology*: Coronal brain slices (thickness ~ 350 μm) from C57Bl6/J mice were prepared using a VT1200S vibrating blade microtome (Leica Biosystems Inc.) in ice-chilled artificial cerebrospinal fluid (aCSF). The temperature of the slices was then elevated to room temperature for > 45 min before starting the experiment with continuous oxygenation (95% $O_2$, 5% $CO_2$). The brain slices were placed in a custom-designed setup with a perfusion chamber containing carbonated aCSF. Baseline recordings of extracellular responses were conducted using a commercial neural amplifier acquired at 30 kHz sampling rate (Intan RHD2000A) to verify setup configuration and neural activity. A Pt/Ir concentric probe was used to stimulate the cortical region using biphasic pulses from a neural stimulator (Grapevine Scout with micro2+stim headstage, Ripple Neuro) and a monopolar tungsten electrode was used for recording.

**References**

1. Machado, T. A., Kauvar, I. v. & Deisseroth, K. Multiregion neuronal activity: the forest and the trees. *Nat Rev Neurosci* **23**, 683–704 (2022).

2. Montelisciani, L., Dijkema, E. & Olcese, U. Single-Neuron and Population Methods to Study the Circuit-Level Cortical Mechanisms of Multisensory Processing. in 1–37 (2025). doi:10.1007/978-1-0716-4208-5_1.

3. Saito, Y., Osako, Y. & Murayama, M. Unraveling the neural code: analysis of large-scale two-photon microscopy data. *Microscopy* **74**, 146–163 (2025).

4. Teichert, T. *et al.* Volumetric mesoscopic electrophysiology: a new imaging modality for the nonhuman primate. *J Neurophysiol* **133**, 1034–1053 (2025).

5. Chen, H. & Fang, Y. Recent developments in implantable neural probe technologies. *MRS Bull* **48**, 484–494 (2023).

6. Perna, A., Angotzi, G. N., Berdondini, L. & Ribeiro, J. F. Advancing the interfacing performances of chronically implantable neural probes in the era of CMOS neuroelectronics. *Front Neurosci* **17**, (2023).

7. Melin, M. D. *et al.* Large scale, simultaneous, chronic neural recordings from multiple brain areas. Preprint at https://doi.org/10.1101/2023.12.22.572441 (2023).

8. Steinmetz, N. A., Zatka-Haas, P., Carandini, M. & Harris, K. D. Distributed coding of choice, action and engagement across the mouse brain. *Nature* **576**, 266–273 (2019).

9. Stringer, C. *et al.* Spontaneous behaviors drive multidimensional, brainwide activity. *Science (1979)* **364**, (2019).

10. Durand, S. *et al.* Acute head-fixed recordings in awake mice with multiple Neuropixels probes. *Nat Protoc* **18**, 424–457 (2023).

11. van Daal, R. J. J. *et al.* Implantation of Neuropixels probes for chronic recording of neuronal activity in freely behaving mice and rats. *Nat Protoc* **16**, 3322–3347 (2021).

12. Ghestem, A. *et al.* Long-term near-continuous recording with Neuropixels probes in healthy and epileptic rats. *J Neural Eng* **20**, 046003 (2023).

13. Bimbard, C. *et al.* An adaptable, reusable, and light implant for chronic Neuropixels probes. *Elife* **13**, (2025).

14. Ferreira-Fernandes, E. *et al.* In vivo recordings in freely behaving mice using independent silicon probes targeting multiple brain regions. *Front Neural Circuits* **17**, (2023).

15. Horan, M. *et al.* Repix: reliable, reusable, versatile chronic Neuropixels implants using minimal components. Preprint at https://doi.org/10.7554/eLife.98977.1 (2024).

16. Cambridge NeuroTech. *Product Catalog*. (2025).

17. Plexon. *N-Form Multi-Electrode Array*. (2025).

18. Rossetti, N., Hagler, J., Kateb, P. & Cicoira, F. Neural and electromyography PEDOT electrodes for invasive stimulation and recording. *J Mater Chem C Mater* **9**, 7243–7263 (2021).

19. Malekoshoaraie, M. H. *et al.* Fully flexible implantable neural probes for electrophysiology recording and controlled neurochemical modulation. *Microsyst Nanoeng* **10**, 91 (2024).

20. Fan, B., Wolfrum, B. & Robinson, J. T. Impedance scaling for gold and platinum microelectrodes. *J Neural Eng* **18**, 056025 (2021).

21. Frampton, J. P., Hynd, M. R., Shuler, M. L. & Shain, W. Effects of Glial Cells on Electrode Impedance Recorded from Neural Prosthetic Devices In Vitro. *Ann Biomed Eng* **38**, 1031–1047 (2010).

22. Salatino, J. W., Ludwig, K. A., Kozai, T. D. Y. & Purcell, E. K. Glial responses to implanted electrodes in the brain. *Nat Biomed Eng* **1**, 862–877 (2017).

23. Sillay, K. A. *et al.* Long-Term Surface Electrode Impedance Recordings Associated with Gliosis for a Closed-Loop Neurostimulation Device. *Ann Neurosci* **25**, 289–298 (2018).

24. Oldroyd, P., Gurke, J. & Malliaras, G. G. Stability of Thin Film Neuromodulation Electrodes under Accelerated Aging Conditions. *Adv Funct Mater* **33**, (2023).

25. Weiss, J. M., Flesher, S. N., Franklin, R., Collinger, J. L. & Gaunt, R. A. Artifact-free recordings in human bidirectional brain–computer interfaces. *J Neural Eng* **16**, 016002 (2019).

26. Zhou, A., Johnson, B. C. & Muller, R. Toward true closed-loop neuromodulation: artifact-free recording during stimulation. *Curr Opin Neurobiol* **50**, 119–127 (2018).

27. Chen, J. *et al.* Recent Trends and Future Prospects of Neural Recording Circuits and Systems: A Tutorial Brief. *IEEE Transactions on Circuits and Systems II: Express Briefs* **69**, 2654–2660 (2022).

28. Steinmetz, N. A., Koch, C., Harris, K. D. & Carandini, M. Challenges and opportunities for large-scale electrophysiology with Neuropixels probes. *Curr Opin Neurobiol* **50**, 92–100 (2018).

29. Steinmetz, N. A. *et al.* Neuropixels 2.0: A miniaturized high-density probe for stable, long-term brain recordings. *Science (1979)* **372**, (2021).

30. Angotzi, G. N. *et al.* SiNAPS: An implantable active pixel sensor CMOS-probe for simultaneous large-scale neural recordings. *Biosens Bioelectron* **126**, 355–364 (2019).

31. Raducanu, B. C. *et al.* Time multiplexed active neural probe with 678 parallel recording sites. in *2016 46th European Solid-State Device Research Conference (ESSDERC)* 385–388 (IEEE, 2016). doi:10.1109/ESSDERC.2016.7599667.

32. Guha, K. K. S. S. I. J. *Micro and Nanoelectronics Devices, Circuits and Systems*. vol. 1382 (Springer Nature, Singapore, 2025).

33. Kinget, P. R. Scaling analog circuits into deep nanoscale CMOS: Obstacles and ways to overcome them. in *2015 IEEE Custom Integrated Circuits Conference (CICC)* 1–8 (IEEE, 2015). doi:10.1109/CICC.2015.7338394.

34. Minkenberg, C., Krishnaswamy, R., Zilkie, A. & Nelson, D. Co-packaged datacenter optics: Opportunities and challenges. *IET Optoelectronics* **15**, 77–91 (2021).

35. Butt, M. A., Janaszek, B. & Piramidowicz, R. Lighting the way forward: The bright future of photonic integrated circuits. *Sensors International* **6**, 100326 (2025).

36. Kim, M. *et al.* O-Band Silicon Ring Modulators With Highly Efficient Electro- and Thermo-Optic Modulation. *Journal of Lightwave Technology* **43**, 1328–1334 (2025).

37. Atabaki, A. H., Eftekhar, A. A., Yegnanarayanan, S. & Adibi, A. Sub-100-nanosecond thermal reconfiguration of silicon photonic devices. *Opt Express* **21**, 15706 (2013).

38. Edinger, P. *et al.* Vacuum-sealed silicon photonic MEMS tunable ring resonator with an independent control over coupling and phase. *Opt Express* **31**, 6540 (2023).

39. Errando-Herranz, C. *et al.* MEMS for Photonic Integrated Circuits. *IEEE Journal of Selected Topics in Quantum Electronics* **26**, 1–16 (2020).

40. *RHD2000 Series Digital Electrophysiology Interface Chips*. (2012).

41. *Neuro Amp EX*. (2024).

42. Ye, Z. *et al.* Ultra-high density electrodes improve detection, yield, and cell type identification in neuronal recordings. Preprint at https://doi.org/10.1101/2023.08.23.554527 (2023).

43. Beau, M. *et al.* A deep learning strategy to identify cell types across species from high-density extracellular recordings. *Cell* **188**, 2218-2234.e22 (2025).

44. Tian, Y. *et al.* Reconfigurable Electro-optic Logic Circuits Using Microring Resonator-Based Optical Switch Array. *IEEE Photonics J* **8**, 1–8 (2016).

45. Ohno, S., Tang, R., Toprasertpong, K., Takagi, S. & Takenaka, M. Si Microring Resonator Crossbar Array for On-Chip Inference and Training of the Optical Neural Network. *ACS Photonics* **9**, 2614–2622 (2022).

46. Posadas, A. B. *et al.* Electro-Optic Barium Titanate Modulators on Silicon Photonics Platform. in *2023 IEEE Silicon Photonics Conference (SiPhotonics)* 1–2 (IEEE, 2023). doi:10.1109/SiPhotonics55903.2023.10141930.

47. Hsu, W.-C., Zhen, C. & Wang, A. X. Electrically Tunable High-Quality Factor Silicon Microring Resonator Gated by High Mobility Conductive Oxide. *ACS Photonics* **8**, 1933–1936 (2021).

48. Chu, H. M. & Hane, K. A Wide-Tuning Silicon Ring-Resonator Composed of Coupled Freestanding Waveguides. *IEEE Photonics Technology Letters* **26**, 1411–1413 (2014).

49. Ikeda, T. & Hane, K. A microelectromechanically tunable microring resonator composed of freestanding silicon photonic waveguide couplers. *Appl Phys Lett* **102**, (2013).

50. Takahashi, K., Kanamori, Y., Kokubun, Y. & Hane, K. A wavelength-selective add-drop switch using silicon microring resonator with a submicron-comb electrostatic actuator. *Opt Express* **16**, 14421 (2008).

51. Sattari, H. *et al.* Silicon photonic microelectromechanical systems add-drop ring resonator in a foundry process. *Journal of Optical Microsystems* **2**, (2022).

52. Liu, K., Sun, S., Majumdar, A. & Sorger, V. J. Fundamental Scaling Laws in Nanophotonics. *Sci Rep* **6**, 37419 (2016).

53. Purcell, E. M. Spontaneous Emission Probabilities at Radio Frequencies. *Physics Reviews* **69**, 681 (1995).

54. Shah Hosseini, E., Yegnanarayanan, S., Atabaki, A. H., Soltani, M. & Adibi, A. Systematic design and fabrication of high-Q single-mode pulley-coupled planar silicon nitride microdisk resonators at visible wavelengths. *Opt Express* **18**, 2127 (2010).

55. Xiao, S., Khan, M. H., Shen, H. & Qi, M. Compact silicon microring resonators with ultra-low propagation loss in the C band. *Opt Express* **15**, 14467 (2007).

56. Okatani, T., Sato, Y., Imai, K., Hane, K. & Kanamori, Y. Improvement of silicon microdisk resonators with movable waveguides by hydrogen annealing treatment. *Journal of Vacuum Science & Technology B, Nanotechnology and Microelectronics: Materials, Processing, Measurement, and Phenomena* **39**, (2021).

57. Roberts, S., Ji, X., Cardenas, J., Corato-Zanarella, M. & Lipson, M. Measurements and Modeling of Atomic-Scale Sidewall Roughness and Losses in Integrated Photonic Devices. *Adv Opt Mater* **10**, (2022).

58. Andersen, Per. *The Hippocampus Book*. (Oxford University Press, 2007).

59. Traub, R. D. & Wong, R. K. S. Cellular Mechanism of Neuronal Synchronization in Epilepsy. *Science (1979)* **216**, 745–747 (1982).

60. Buzsáki, G. & Draguhn, A. Neuronal Oscillations in Cortical Networks. *Science (1979)* **304**, 1926–1929 (2004).

61. Buzsáki, G., Anastassiou, C. A. & Koch, C. The origin of extracellular fields and currents — EEG, ECoG, LFP and spikes. *Nat Rev Neurosci* **13**, 407–420 (2012).

**Acknowledgements**

The authors would like to thank Zabir Ahmed for his assistance in development of the photonics fabrication process and experimental setup. This work has received funding from the National Science Foundation (Grant No. 2111660), National Institutes of Health (Grant No. 5R21EY033084-02) (Corresponding authors: Maysamreza Chamanzar and Gianluca

Piazza), and National Science Foundation Graduate Research Fellowship Program (Grant No. DGE2140739) (Corresponding author: Andrew Cochran). Any opinions, findings, and conclusions or recommendations expressed in this material are those of the author(s) and do not necessarily reflect the views of the funding organizations. The authors acknowledge the use of the Bertucci Nanotechnology Laboratory and Materials Characterization Facility at Carnegie Mellon University supported by grants BNL-78657879 and MCF-677785.

**Conflict of Interest Statement**

The authors declare no conflicts of interest.